\documentclass[floatfix,showpacs,amsmath,amssymb,letterpaper,groupaddresses,superscriptaddress]{article}
\usepackage{graphicx}
\usepackage{latexsym}
\usepackage{amsmath}
\usepackage{amssymb}
\usepackage{epsfig}
\usepackage{graphicx}
\usepackage{graphicx}
\usepackage{amssymb,amsmath,times}
\usepackage{hyperref}
\usepackage{color}

\def\a{\alpha}

\def\be{\begin{equation}}
\def\ee{\end{equation}}
\def\ba{\begin{eqnarray}}
\def\ea{\end{eqnarray}}
\def\la{\langle}
\def\ra{\rangle}
\def\a{\alpha}

\def\sq{\square}

\def\h{\hskip 1cm}

\def\lo{\longrightarrow}

\usepackage{epsfig}
\usepackage{graphicx}
\begin{document}
\begin{center}
{\Large \bf An algorithmic proof for the completeness of two-dimensional Ising model}\\
\vspace{1cm} Vahid Karimipour\footnote{email: vahid@sharif.edu}\h
Mohammad Hossein Zarei\footnote{email:mhzarei@physics.sharif.edu}
\vspace{5mm}

\vspace{1cm} Department of Physics, Sharif University of Technology,\\
P.O. Box 11155-9161,\\ Tehran, Iran
\end{center}
\vskip 3cm

\begin{abstract}
We show that the two dimensional Ising model is complete, in the sense that the partition function of any lattice model on any graph is equal to the partition function of the 2D Ising model with complex coupling. The latter model has all its spin-spin coupling equal to $i\frac{\pi}{4}$ and all the parameters of the original model are contained in the local magnetic fields of the Ising model. This result has already been derived by using techniques from quantum information theory and by exploiting the universality of cluster states. Here we do not use the quantum formalism and hence make the completeness result accessible to a wide audience. Furthermore our method has the advantage of being algorithmic in nature so that by following a set of simple graphical transformations, one is able to transform any discrete lattice model to an Ising model defined on a (polynomially) larger 2D lattice.

\end{abstract}
\section{Introduction}

Despite the fact that foundations of statistical mechanics were laid down in the nineteen century, the understanding that a single partition function can describe
different phases of matter, is rather recent and indeed as late
as 1930's, there was not a consensus among physicists that one single
partition function can describe ice, water and vapor \cite{Golden} or can give a sharp phase transition which we observe in ferromagnetic materials. The works of
Kramers, Wannier, Onsager \cite{onsager, kramers, kaufman} and others gradually
established beyond doubt that in the thermodynamic limit,
singular behavior and phase transition can arise from a single
partition function based on a single model Hamiltonian. For
example the Ising Hamiltonian can describe both the ordered phase
of a ferromagnet and the disordered phase of a paramagnet. However, the study of second order phase transition over the past decades, revealed an interesting feature which is called universality. It means that the near
the point of second order phase transition, the details of
the model Hamiltonian do not matter and only some general
properties
like symmetries and dimension of the order parameter are important. \\

Our current understanding of this important phenomena, that is the independence of the most salient features of phase transition from the microscopic details is based on Renormalization Group (RG) and effective theories. Under renormalization group (change of scale), different microscopic theories tend toward an effective theory which describes the long range characteristics of a whole class of models, irrespective of their microscopic details. Therefore if we imagine an infinite dimensional space in which all forms of Hamiltonians describing the interactions of Ising type variables are defined, there are some special points in this space, to which other models tend toward the RG flow. These fixed points essentially capture the most important properties of phase transitions of all the models along the RG trajectories.\\

Quite recently a new notation, different from universality,  has emerged in the study of classical statistical mechanical models which is appropriately called completeness \cite{completeness, delgado, u1, phi,statistic, nest}. The idea of completeness asserts that certain statistical mechanical models are complete in the sense that if the partition function of these complete models are solved,  then the partition function of any other model can be also solved with a polynomial computational overhead. This new insight certainly is different from the notion of universality which applies only to the long distance limit of different theories. It takes into account the microscopic details of different models and yet claims that all these models are inherited from one single model. It is reminiscent of the concept of complete problems in computability theory \cite{np, np2}. Therefore, in the same space of coupling constants mentioned above, there are certain special points, from which a whole class of other models can be derived, in the sense that the partition function of the latter is exactly equal to the partition function of the complete model if we formulate the complete model on a sufficiently large lattice. \\

Up until now many models have been shown to be complete \cite{completeness, delgado, newj, u1, phi}. In this paper we are concerned with the Ising model which was the first to be proved complete \cite{completeness}. It is defined on a 2D rectangular lattice, with the following Hamiltonian

\begin{equation}
\beta H=\sum_k h_i S_i +i\frac{\pi}{4}\sum_{\la i, j\ra}S_iS_j,
\end{equation}
where $S_i=\pm 1$, the $h_i$'s are local magnetic fields which can be complex and all the interactions are nearest neighbors. \\

It is true that the model has no physical realization due to the complex fields and couplings, however it is quite significant that any other model, regardless of the nature of its statistical variables, dimensionality and even the geometry of its lattice, and the nature of its interactions is just an instant of this general computational problem.
All the details of the original model, the couplings, the magnetic fields, the connectivity and geometry of the lattice are contained in the size of the 2D lattice and the inhomogeneous complex magnetic fields. \\

Almost all of these completeness results
 \cite{completeness,delgado,newj,u1,phi,statistic,nest} were derived by merging of ideas from
statistical mechanics and quantum information theory \cite{nest2, other, vidal, lidar, lidar1, lidar2, lidar3, lidar4, lidar5, lidar6, lidar7}.The new paradigm of Measurement-based Quantum Computation (MQC)
and the universality of cluster states for
MQC \cite{mqc4, dmqc1, dmqc2, dmqc3, dmqc4, dmqc5, dmqc6, dmqc7, dmqc8, dmqc9, mqc, mqc2, mqc3, mqc5, mqc6, mqc7} was essential in these derivations.\\
The starting point was the observation in \cite{quantum formalism} that the
partition function of any given discrete model can be written as
a scalar product,
\begin{equation}\label{par1}
  Z_G (J)=\la \a |\Psi_G\ra,
\end{equation}
where $\la \a|$ is a product state encoding all the coupling
constants $J$ and $|\Psi_G\ra$ is an entangled {\it graph} state,
 defined on the vertices and edges of a graph, and encoding the geometry of the lattice. The core concept of completeness is the fact that the $2D$ cluster state is universal. More
concretely we know that the graph state $|\Phi_G\ra$
corresponding to a graph $G$ can be obtained from an appropriate
cluster state $|\Psi_\square\ra$ corresponding to a rectangular
lattice (denoted by $\square$), through a set of adaptive
single-qubit measurements ${\cal M}$. The measurements being
single-qubit can be formally written as $|\a_M\ra\la \a_M|$,
where $|\a_M\ra$ is a product state encoding  the bases
and the results which have been measured. Thus the totality of
measurements ${\cal M}$ transforms the cluster state as follows
\begin{equation}\label{measure}
  |\Psi_\square\ra\lo |\a_M\ra\la a_M|\Psi_\square\ra=|\a_M\ra\otimes |\Psi_G\ra,
\end{equation}
which shows that after disregarding the states of the measured
qubits $|\a_M\ra$, what is left is an appropriate graph state
\begin{equation}\label{relation1}
|\Psi_G\ra=\la \a_M|\Psi_\sq\ra.
\end{equation}
Combination of this relation with (\ref{par1}), leads to the
completeness result mentioned above. That is one writes
\begin{equation}\label{relation}
Z_G(J)\equiv \la \a |\Psi_G\ra=\la \a , \a_M|\Psi_\sq\ra,
\end{equation}
and notes that $\la \a, \a_M|$ now encodes a set of generally
inhomogeneous pattern of interactions on the cluster state.
Therefore one has
\begin{equation}\label{relation}
Z_G(J)\equiv Z_\sq(J,J').
\end{equation}

However, there are two shortcomings in the above derivation. First it is an existence rather than a constructive proof, and one does not learn how to reduce any given statistical model to the 2D Ising model, at least not in a simple way which we intend to introduce. Second, it is not accessible to those who are not familiar with the concepts and techniques of quantum information theory, notably that of measurement-based quantum computation. While the result is a general one  in statistical mechanics with potential intersect across many disciplines, its derivation is somewhat specialized. \\

It is therefore the aim of our paper to provide a  proof for this completeness result which is both constructive and accessible to a large audience and readership. We show how any model can be reduced to an instance of 2D Ising model through a set of simple graphical transformation steps. As a by-product we also re-derive the well-known duality relations between various Ising models, in a way which is different from the high and low temperature expansions. Finally our proof does not require any basic knowledge of quantum information theory or MQC or even quantum mechanics. We hope that by providing this proof, we will make this important result accessible to a wide audience who can further explore the far-reaching consequence of the completeness of 2D Ising model. \\

The structure of this paper is as follows: In section (\ref{sec1}), we first review the basic result that any discrete statistical mechanical model can be expressed in terms of Ising variables. We then show how the interaction terms of any such model can be transformed to the local magnetic fields of an Ising model whose interaction strength is fixed at the value of $i\frac{\pi}{4}$.  In section (\ref{sec2}) we apply this idea to a few well-known examples and use it to derive the well-known results on duality in a way, which is independent from the high and low temperature expansions. In section (\ref{com}) we come to the main result of our paper, which is the proof of completeness of 2D Ising model. We introduce basic transformation rules and express them in terms of diagrams (with further examples in an appendix) which allows us to transform any Ising model on any kind of graph (planar or not), to the 2D Ising model on rectangular lattices. The paper ends with a discussion and the above mentioned appendix. \\

\section{Transformation in lattice models}\label{sec1}
Consider an arbitrary discrete lattice model, like Ising, Potts, vertex or face model \cite{Baxter}. The discrete variables of this lattice model need not be bi-valued like the Ising model, and the lattice need not be regular.  It has been shown in \cite{delgado} that the Hamiltonian of such a model can always be re-written in terms of two-valued spin variables , in the form
\begin{equation}\label{eq1}
H=\sum J_{i}S_{i}+\sum J_{ij}S_{i}S_{j}+\sum J_{ijk}S_{i}S_{j}S_{k}+... ,
\end{equation}
where $S_{i}=\{1,-1\}$ is a spin variable and $J_{i}, J_{ij},..$ are coupling constants. Therefore the price that one pays for converting any type of variable to the Ising variable is to admit few-body interactions in the Hamiltonian in addition to two-body interactions. As two simple examples, consider the the three-state and four- state Potts model, whose variables $q$ take 3 or 4 different variables respectively, and their Hamiltonians are given by:
\begin{equation}
H=-J\sum_{\la i,j\ra} \delta_{q_i,q_j}.
\end{equation}
 As shown in appendix A, by suitable encodings of the Potts variables $q_i$ into a a pair of Ising variables $(S_i,S'_i)$, we can turn these models into Ising models on lattices of 2N sites.\\

These two examples, illustrate the general idea of (\ref{eq1}) that by encoding any kind of discrete variables into a number of Ising variables, one can turn any statistical model into an Ising model, albeit with complicated multi-particle interactions.\\

Having this result, it is now enough to consider Ising model on arbitrary graphs with arbitrary patterns of interactions. The Hamiltonian of such a model is written as
\begin{equation}
H= \sum J_{i}S_{i}+\sum J_{ij}S_{i}S_{j}+\sum J_{ijk}S_{i}S_{j}S_{k}+\cdots,
\end{equation}
where $J_i$, $J_{ij}$ and $J_{ijk}$ are the strengths of one, two and three body interactions and $\cdots$ stand for interactions of higher number of spins.
We can now turn this Hamiltonian into an equivalent one, containing only on-site interactions or local magnetic fields. The procedure is to define new spin variables as
\begin{equation}\label{ee1}
S_{ij}=S_{i}S_{j}~~~~,~~~~S_{ijk}=S_{i}S_{j}S_{k}, \ \ \cdots,
\end{equation}
which turn the above Hamiltonian into
\begin{equation}
H= \sum J_{i}S_{i}+\sum J_{ij}S_{ij}+\sum J_{ijk}S_{ijk}+\cdots.
\end{equation}
Note that the variables $S_{ij}$ are assigned to links, the variables $S_{ijk}$ to triangles and other variables to higher simplexes of the lattice. \\

If all these new variables were independent, then the new Hamiltonian would have been that of a collection of spins $\bar{S}_\a \in \{S_i, S_{ij}, S_{ijk},\cdots\} $ in local magnetic fields $h_\a\in  \{J_i, J_{ij}, J_{ijk}\}$ with a trivial partition function. However, the price that we should pay for this change of variable is that the new variables are not independent. In fact as we see, this price has the extra merit of paving the way for the proof of completeness result. There are many constraints between the new variables, for example we have
\begin{equation}
S_{i}S_{ij}S_{j}=1~~~,~~~S_{i}S_{jk}S_{ijk}=1, \ \  \cdots.
\end{equation}
In order to calculate the partition function in a correct way, one should take into account all these constraints. Therefore the general form of the partition function will be
\begin{equation}
\mathcal{Z}=N\sum_{\{\bar{S}_\a\}} e^{-\beta \sum h_{\a}\bar{S}_{\a}} \prod_{C}  \delta(\prod_{\a\in C} \bar{S}_\a, 1).
\end{equation}
In the above formula, $C$ denote a constraint and $\a \in C$ stands for the spin variables taking part in a constraint $C$, and $N$ is a numerical factor which takes of overcounting. For simplicity of notation in what follows we denote the new variables $\bar{S}$ with the ordinary variable $S$.\\
The basic idea is now to rewrite the constraints, which contain product of spins,  in such a way to turn them into a form containing only two-body interactions, no matter how complicated the constraint is. To do this we introduce the following lemma:\\

$\mathbf{Lemma:}$ For each constraint, we can add one single spin variable $S_0$ and write it as
\begin{equation}\label{L1}
\delta(S_1S_2\cdots S_m ,1)=\frac{1}{2}\sum_{S_0=1,-1}e^{\frac{i\pi}{4}(1-S_0)(m-S_1-S_2-\cdots S_m)}.
\end{equation}
When inserted into the partition function, this produces extra magnetic fields $-i\frac{\pi}{4}$ on each spins $S_i$, a magnetic field of $-m\frac{i\pi}{4}$ on the extra spin $S_0$ and two-body interactions of strength $\frac{i\pi}{4}$ between pairs of spins. The proof of lemma is very simple:\\

\begin{figure}[t]
\centering
\includegraphics[width=7cm,height=3.5cm,angle=0]{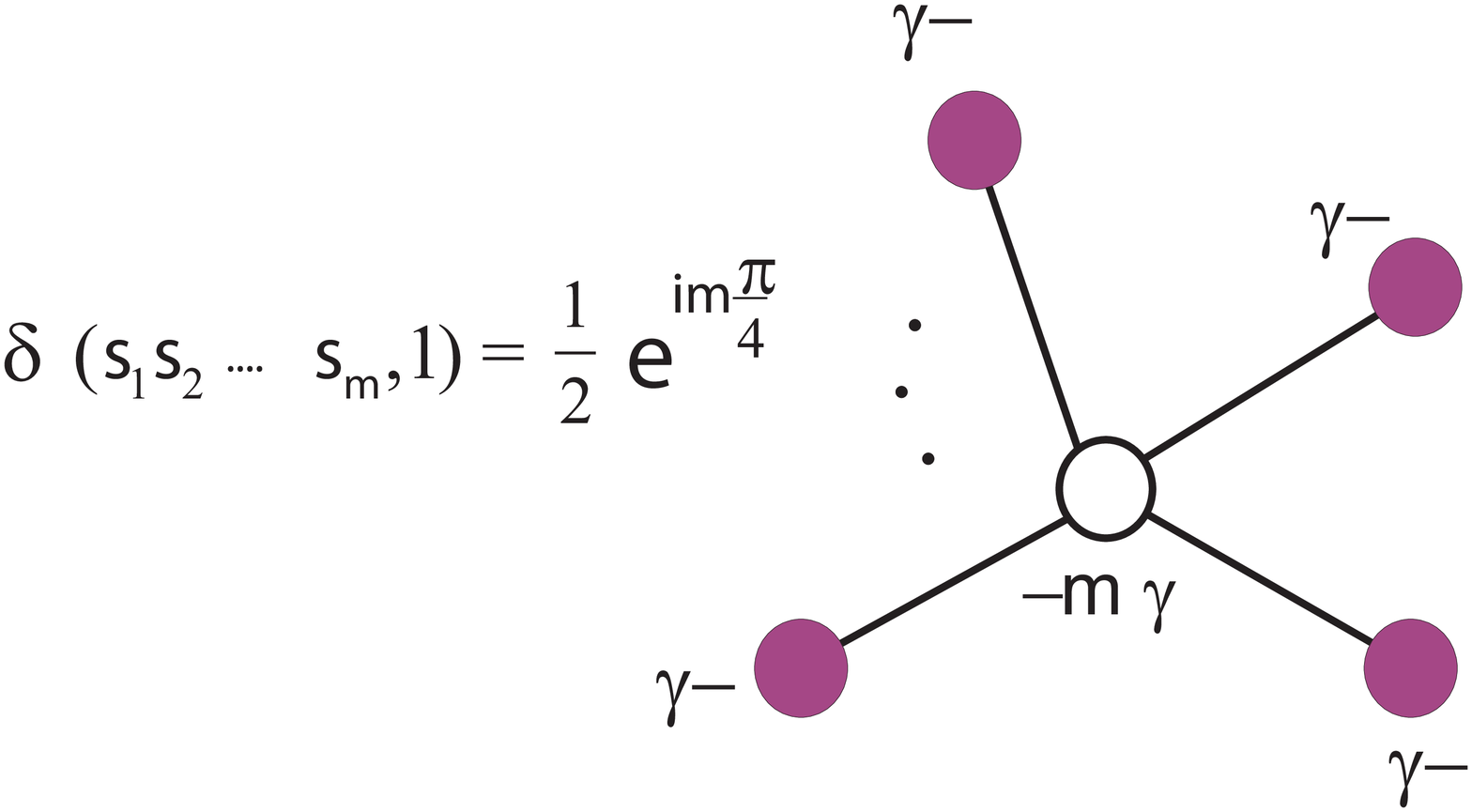}
\caption{(Color Online) Graphical representation of lemma (\ref{L1}). A white circle denotes a spin which is summed over. The values on spins indicate the local magnetic fields. Any link indicates a coupling with value $\frac{i\pi}{4}$. $ m$ is the number of black spins.
} \label{fig0}
\end{figure}

$\mathbf{Proof:}$ We use the variables $\sigma\in \{0,1\}$ defined as $S=(-1)^{\sigma}=e^{i\pi\sigma}$ to write the constraint as 
\begin{equation}
\delta(S_1S_2\cdots S_m,1)=\delta(\sigma_1+\sigma_2+\sigma_m , 0) = \frac{1}{2}\sum_{\sigma_0=0,1}e^{i\pi\sigma_0(\sigma_1+\sigma_2+\cdots +\sigma_m)}.
\end{equation}
Now going  to the original Ising variables through the relation $\sigma=\frac{1-S}{2}$, converts this last relation to equation (\ref{L1}), hence the lemma is proved.
Lemma (\ref{L1}) is depicted graphically in figure (\ref{fig0}). For this and the other identities which are depicted in other figures, we will use a convention which we gather as follows:\\

\textbf{Notations and Conventions:}
 In all the equations and figures in this section, the Boltzman factor $-\beta=\frac{-1}{k_BT}$ has been absorbed in the Hamiltonian, so $H$ means $-\beta H$ in all the equations.  The parameter $\gamma$ has the fixed value of $i\frac{\pi}{4}$. A spin which is summed over, is depicted by a white circle. A gray circle means a spin whose value has been fixed to $S=1$. Every link indicates an interaction with a coupling strength of $i\frac{\pi}{4}$ between its endpoints. The value written near any white circle is the value of the local magnetic field on this spin.  When a $+$ or $-$ sign comes on the right hand side of a numerical value on spin,  it means that the effect of summation over the white spin is, the addition or subtraction of that value to the local magnetic field of that spin.  \\

Using this identity, the partition function of any model will be equivalent to the partition function of an Ising model on a complex graph in which the coupling constants of all bilinear couplings is the constant value of $i\frac{\pi}{4}$. It is clear that for any model which contain interactions of 2-body, 3-body up to a finite $k-body$ interactions, the number of constraints is polynomially dependent on the number of vertices of the original graph and hence the size of the final Ising graph is only polynomially larger than the original graph. Moreover our construction shows a specific algorithm for turning the original discrete model to an Ising model with only $i\frac{\pi}{4}$ interactions.  We still have to show that the new Ising model on the complex graph which may well be non-planar, is equivalent to a 2D Ising model. We will prove this completeness result in section (\ref{com}). However before doing this, it is instructive to present a few examples to show how the above transformation maps some well known models to other ones.

 \section{Examples}\label{sec2}
 In this section we consider a few examples to show how the transformation (\ref{L1}), makes a discrete model equivalent to an Ising model on a different lattice. The first two examples deal with Ising model with two-body interactions on triangular and hexagonal models respectively,while the third one deals with a new model introduced recently in \cite{three, three2} with three body interactions.\\
 \subsection{Ising model on triangular lattice}
 Consider an Ising model without magnetic field on a triangular lattice with free boundary conditions.  The partition function of this model is
 \begin{equation}
 \mathcal{Z}_\Delta(J)=\sum_{\{S_{i}\}}e^{ J\sum_{\la i,j\ra}S_{i}S_{j}}.
 \end{equation}
\begin{figure}[t]
\centering
\includegraphics[width=5cm,height=5cm,angle=0]{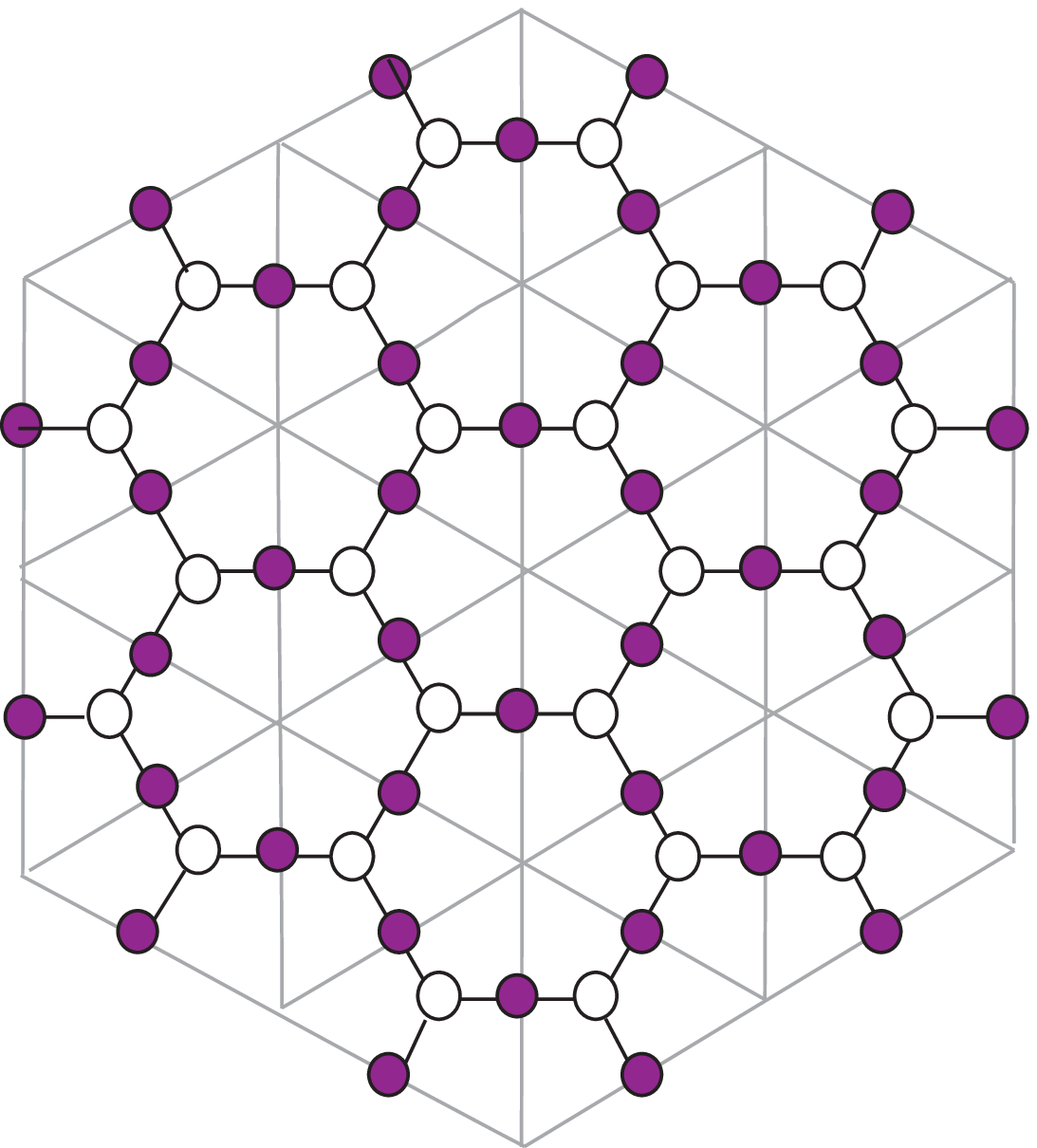}
\caption{(Color Online) An Ising model on a triangular lattice. On each link of the lattice, we put a new spin $S_{ij}:=S_iS_j$. These new spins are constrained, in that the product of all of them around a triangle is equal to 1. This is accounted for by inserting a white spin inside each triangle and using lemma (\ref{L1}). The result is a hexagonal bi-partite lattice. Summing over the black spins will lead to duality between Ising models on triangular and hexagonal lattices. See subsection (\ref{SD}).
} \label{fig}
\end{figure}
 On each edge $e:=(i,j)$ of the lattice, which connects the vertices $i$ and $j$, we insert a new variables $S_{e}=S_{i}S_{j}$ . These new variables are not independent. In fact for each triangle of the lattice with edges $e_1=(i,j)$, $e_2=(j,k)$ and $e_3=(k,i)$, we have $S_{e_1}S_{e_2}S_{e_3}=1$. Implementing this constraint via the relation (\ref{L1}) or the corresponding figure (\ref{fig0}), we arrive at a bi-partite hexagonal lattice, shown in figure (\ref{fig}),  with two kind of vertices, which we designate as black (B) and white (W). The black circles designate the spins $S_e$ put on the edges of the original lattice, and the white circles designate the new spins which take care of the constraint through equation (\ref{L1}). From figure (\ref{fig}), we find that Hamiltonian of the new model is as follows:
 \begin{equation}
H_{hex}=(J-\frac{i\pi}{2})\sum_{i\in B}S_i -\frac{3i\pi}{4}\sum_{j\in WB}S_j+\frac{i\pi}{4}\sum_{i\in B, j\in W}S_iS_j.
\end{equation}
 Taking into account the multiplicative factor for each triangle in figure (\ref{fig}), we arrive at the following relation between partition functions:

 \begin{equation}
Z_\Delta(J)=(\frac{1}{2}e^{3i\frac{\pi}{4}})^{|\Delta|}Z_{hex}(J),
\end{equation}
 where $|\Delta|$ is the number of triangles in the original lattice.

 \subsection{Ising model on hexagonal lattice}

\begin{figure}[t]
\centering
\includegraphics[width=5cm,height=5cm,angle=0]{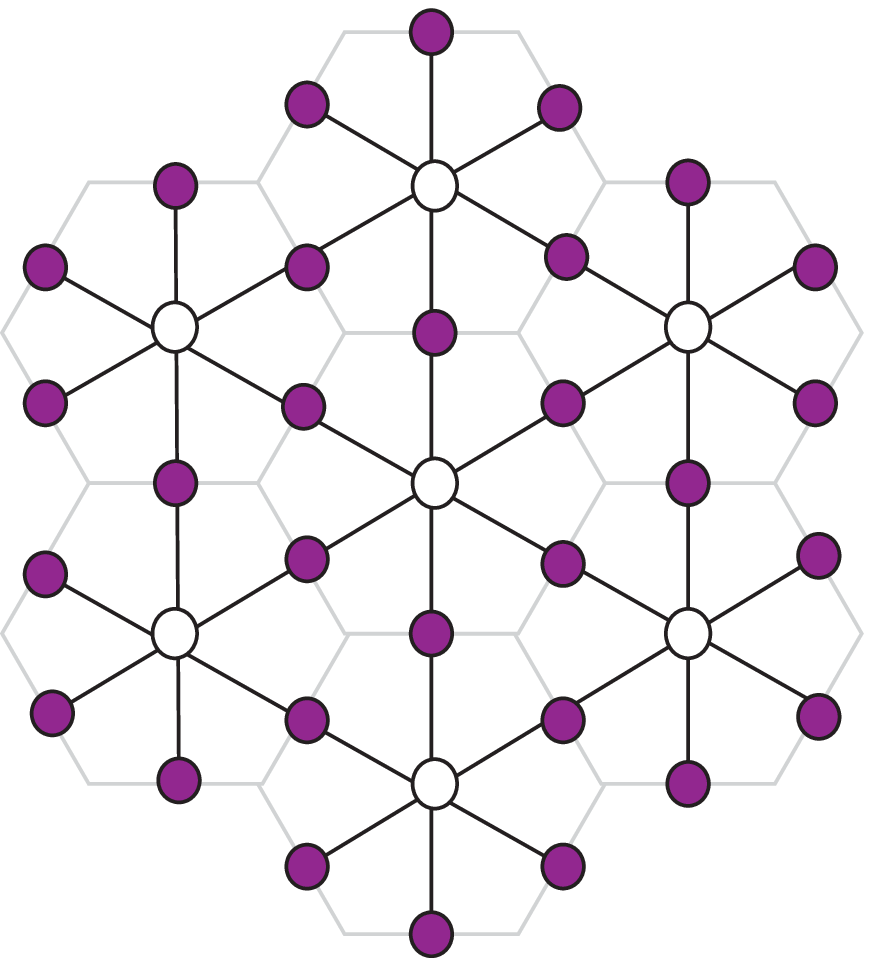}
\caption{(Color Online) An Ising model on a hexagonal lattice. On each link of the lattice, we put a new spin $S_{ij}:=S_iS_j$. These new spins are constrained, in that the product of all of them around a hexagon is equal to 1. This is accounted for by inserting a white spin inside each hexagon and using lemma (\ref{L1}). The result is a triangular bi-partite lattice. Summing over the black spins will lead to duality between Ising models on hexagonal and triangular Ising lattices. See subsection (\ref{SD}).
} \label{fig2}
\end{figure}

 Consider now the Ising model on a hexagonal lattice as shown in figure (\ref{fig2}). The Hamiltonian is

 \begin{equation}\label{H1}
H_{hex}=J\sum_{i\in hex} S_iS_j,
\end{equation}
 where $hex$ means all the vertices of the hexagonal lattice.
  Again we do the same transformation as before, that is on any edge $e=(i,j)$ connecting the two vertices $i$ and $j$ which allocate the spin variables $S_i$ and $S_j$, we put a new spin variable $S_{e}:=S_iS_j$. The original coupling constant $J$  in the Hamiltonian (\ref{H1}) will now play the role of a magnetic field in the new lattice. However the new variables are not independent and indeed on for any hexagon in the lattice, the product of all the spin variables around the edges is equal to 1. Hence using (\ref{L1}) we find that the model is equivalent to a new triangular lattice, as shown in figure (\ref{fig2}). Again the new lattice is bi-partite, with two sets of sites denoted as black (B) and white (W) as in the previous example. The Hamiltonian of the new model is:

    \begin{equation}
H_{\Delta}=(J-i\frac{\pi}{2})\sum_{i\in B} S_i-\frac{3i\pi}{2}\sum_{j\in W}S_j+i\frac{\pi}{4}\sum_{i\in B, j\in W}S_iS_j.
\end{equation}

In view of the relation in figure (\ref{fig2}), the relation of the partition functions will be

   \begin{equation}
Z_{hex}(J)=(\frac{1}{2}e^{3i\frac{\pi}{2}})^{|hex|}Z_{\Delta}(J),
\end{equation}
  where $|hex|$ is the number of hexagons in the original lattice.

 \subsection{An Ising model with three-body interaction}
\begin{figure}[t]
\centering
\includegraphics[width=5cm,height=5cm,angle=0]{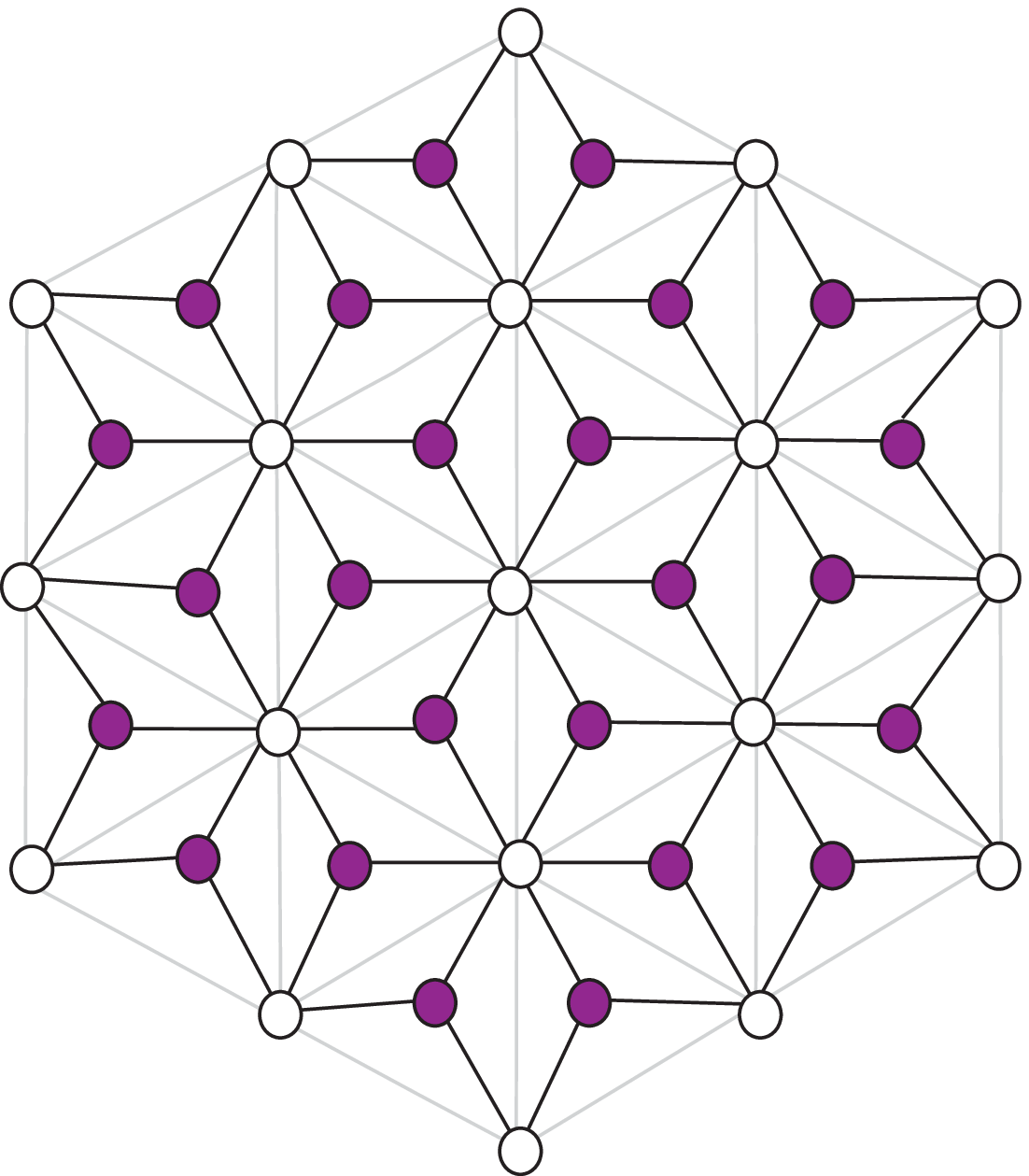}
\caption{(Color Online) An Ising model on a triangular lattice with three-body interactions.  On each face of the lattice, we put a new spin $S_{ijk}:=S_iS_jS_k$. These new spins are constrained, in that the product of all of them around a hexagon is equal to 1. This is accounted for by inserting a white spin inside each hexagon and using lemma (\ref{L1}). The result is a triangular bi-partite lattice.} \label{fig3}
\end{figure}

The final example is a kind of Ising model on triangular lattice with three body interactions. This example was studied in \cite{color} in the context of topological color codes, it is in a different universality class than the usual Ising universality class \cite{three2}. The Hamiltonian of this model is

 \begin{equation}
H^3_{\Delta}=J\sum_{\la i,j,k\ra} S_iS_jS_k,
\end{equation}
  where by $\la i,j,k\ra$, we mean the vertices belonging to a single triangle. In this case, the new spin variables are assigned to each triangle (plaquette), as $S_t:=S_iS_jS_k$, where $i,j, $ and $k$ are the vertices of the triangle $t$. The coupling constant $J$ plays the role of local magnetic field on each plaquette. The new spin variables satisfy the constraint that $\prod'_{t}S_t=1$, where by $\prod'$ we mean the six triangles sharing one single vertex in the center. The new lattice is shown in figure (\ref{fig3}), and is seen to be triangular again, which we denote by $\Delta^*$,  with the following Hamiltonian:

   \begin{equation}
H^3_{\Delta^*}=(J-i\frac{3\pi}{4})\sum_{i\in B} S_i-3i\frac{\pi}{2}\sum_{j\in W}S_j+i\frac{\pi}{4}\sum_{i\in B, j\in W}S_i S_j.
\end{equation}

Taking into account the multiplicative factors in figure (\ref{fig3}), we find the following relations between the partition functions:
   \begin{equation}
Z_\Delta(J)=(\frac{1}{2}e^{3i\frac{\pi}{2}})^{|t|}Z_{\Delta^*}(J),
\end{equation}
where $|t|$ is the number of triangles in the original lattice. Note that the new lattice has only two-body interactions. \\

\subsection{Duality}\label{SD}
As the reader may have noticed, the graphs obtained for the above examples, are all bi-partite graphs. This allows us to do a partial summation over the subgraphs, comprised of black spins and obtain an equivalent Ising model. This then leads to the well-known duality relations. As it is evident from figure (\ref{fig2}), doing such a partial summation on the final Ising graphs which were derived from the hexagonal and triangular Ising models, leaves us with Ising models on triangular and hexagonal lattices respectively, but with different couplings. To explicitly demonstrate this duality, we detail it only for the rectangular lattice which is known to be self-dual.  Repeating the analysis for other kinds of models is straightforward. \\

\begin{figure}[t]
\centering
\includegraphics[width=5cm,height=5cm,angle=0]{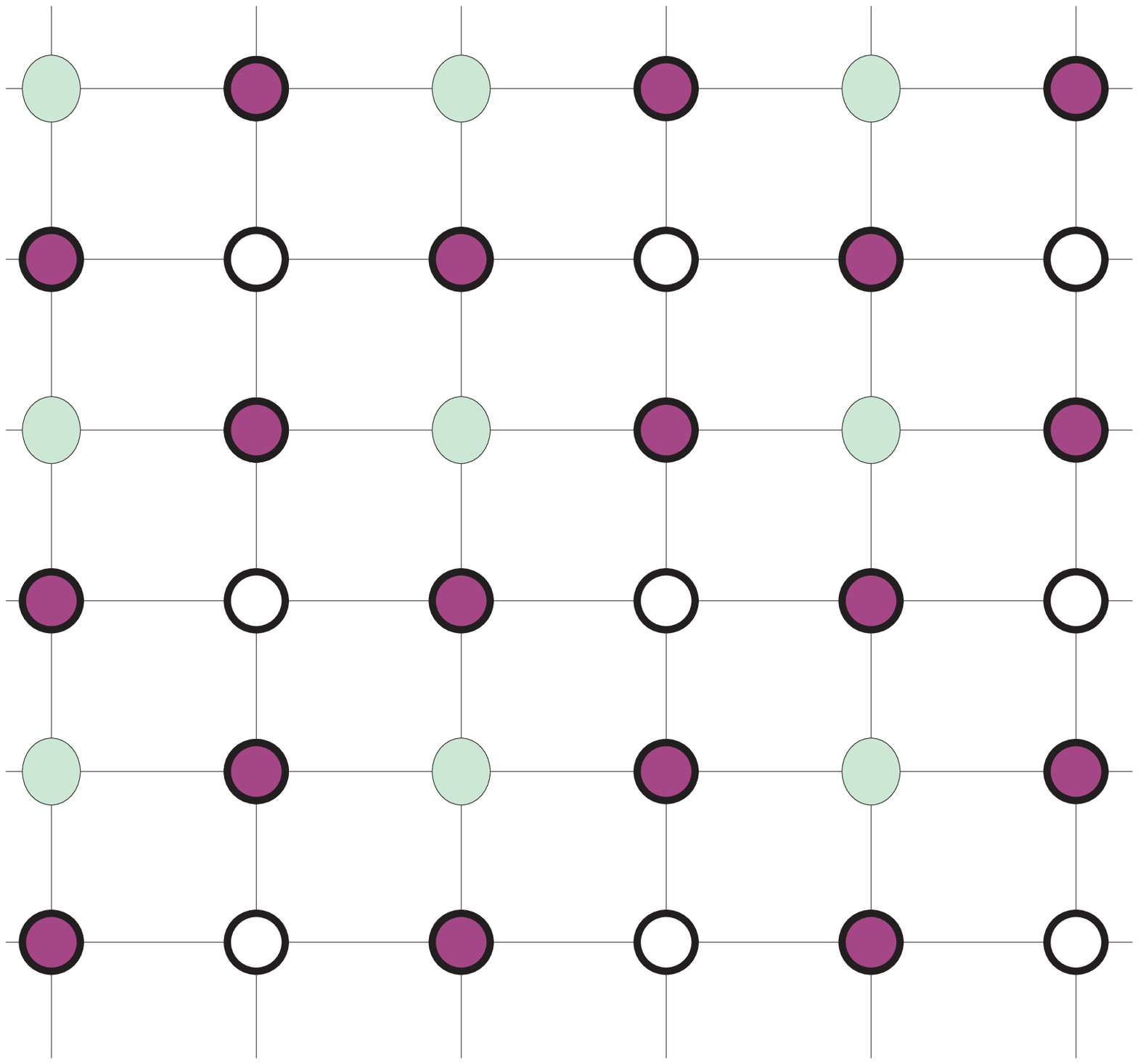}
\caption{(Color Online) The Ising model on a rectangular lattice and its dual. The light-blue circles are the spins on the original lattice, interacting with the couplings $J$. The black spins are the new spins which are assigned to each link as in (\ref{ee1}) and the constraint between these new spins around each plaquette, is taken care of by addition of a single white spin in each plaquette.
} \label{fig31}
\end{figure}

Figure (\ref{fig31}) shows the 2D Ising model. The light-blue circles are the spins on the original lattice, interacting with the couplings $J$. The black spins are the new spins which are assigned to each link as in (\ref{ee1}) and the constraint between these new spins around each plaquette, is taken care of by addition of a single white spin in each plaquette. The new bipartite lattice is governed by the following Hamiltonian:
\begin{equation}\label{22}
\beta H=(J-\frac{i\pi}{2})\sum_{i\in B}S_{i}+i\pi\sum_{j\in W}S_{j}+\frac{i\pi}{4}\sum_{i\in B, j\in \{W\in N_{i}\}}S_{i}S_{j},
\end{equation}
where the following relation exists between the original partition function and the new one:
\begin{equation}
Z_{2D Ising}(J)=e^{iN\pi}Z^*_{2D Ising}
\end{equation}
Here $N$ is the number of spins of the rectangular lattice and $Z^*$ is the partition function of the equivalent model.
If we now sum over the black spins, we obtain a new Ising model comprised entirely of white spins. To find the Hamiltonian of the new spin model, a simple calculation is in order.

 Consider a black spin $S_0$ and it's two neighboring white spins, $S_1$ and $S_2$. Summing this part of the partition function we find
\begin{equation}
\sum_{S_{0}}e^{\frac{i\pi}{4}S_{0}(S_{1}+S_{2})+(J-\frac{i\pi}{2})S_{0}}= 2 \cosh(\frac{i\pi}{4}(S_{1}+S_{2})+(J-\frac{i\pi}{2})).
\end{equation}
After a simple algebra, the right hand side can be rewritten in the form
\begin{equation}
\cosh(\frac{i\pi}{4}(S_{1}+S_{2})+(J-\frac{i\pi}{2}))=Ae^{KS_{1}S_{2}+h_{1}S_{1}+h_{2}S_{2}}
\end{equation}
where the parameters are required to satisfy the following relations:
$$Ae^{K+h_{1}+h_{2}}=\cosh(J),$$
$$Ae^{-K+h_{1}-h_{2}}=-i\sinh(J),$$
$$Ae^{-K-h_{1}+h_{2}}=-i\sinh(J),$$
$$Ae^{K-h_{1}-h_{2}}=-\cosh(J),$$
the solution of which is
$$
h_{1}=h_{2}=\frac{i\pi}{4}$$
$$A^2=-\frac{1}{2}\sinh(2J)$$
\begin{equation}
e^{-2K}=tanh(J).
\end{equation}
The last relation is the well-known duality relation in the 2D Ising model, which is usually obtained by  comparing the high and low temperature expansion of the Ising model partition function. Here it is derived without any reference to high or low temperature expansions. Putting all this together, we find
\begin{equation}
Z_{_{{\rm 2D Ising}}}(J)=2^N \sinh(2J)^N Z_{_{{\rm 2D Ising}}}(K),
\end{equation}
where N is the number of spins of the 2D lattice. This is exactly the duality relation of the 2D Ising model. This type of analysis can be repeated for triangular and hexagonal lattice models with ensuing duality relations between them.
These examples show how the mapping works and how every type of interaction in the original Hamiltonian turns into a magnetic field and how the constraints turn into complex two-body interactions of $i\frac{\pi}{4}$ in the new model.  We have deliberately considered simple examples in which the constraint do not produce complex non-planar graphs. In general this can happen and the new graph can be topologically complex. Nevertheless we show that by a series of simple transformations, the new model can be turned into a 2D Ising model. Therefore we provide an algorithmic approach for transforming any discrete model to the 2D Ising model, and in this way we provide a different proof of completeness of 2D Ising model. Compared to the existence proof \cite{completeness} based on universality of cluster states, this new proof is constructive in nature.

 \section{Completeness of the Ising model on 2D square lattices}\label{com}
 In previous sections we have shown that any discrete statistical mechanical model is equivalent to an Ising model with complex two-body interactions of $\frac{i\pi}{4}$ where the effect of the original coupling constants is transferred to the local magnetic fields on the spins. Depending on the original statistical mechanical model, the graph of the equivalent Ising model may be complicated and usually it may be a non-planar graph. We now introduce the steps by which one can transform such an Ising model to one  on a  polynomially larger two-dimensional rectangular lattice.
 The steps or the rules go with the name of uniformizing and flattening steps respectively. We describe them in that order.  The notations and conventions introduced before, apply to all which follows.

 \subsection{Uniformizing the Graph}

\begin{figure}[t]
\centering
\includegraphics[width=6cm,height=1.8cm,angle=0]{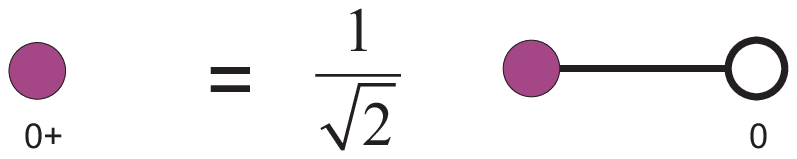}
\caption{(Color Online) Identity (\ref{i1}): the white spin has zero magnetic field and has been summed over. This allows us to increase the degree of any vertex as required. Note that the magnetic field of the black spin does not change, hence the notation $0+$ on this spin.
} \label{UniformUP}
\end{figure}

 As shown in section (\ref{sec1}) and figure (\ref{fig0}), for any constraint between a set of spins $\{S_1, S_2, \cdots S_m\}$ on vertices $\{1, 2, \cdots m\}$,  we have to add a new vertex with spin variable $S$, which links connecting this new vertex to all the vertices $1$ through $m$. This may produce a complex graph with vertices with various degrees of connectivity.  The uniformizing  rule allows us to transform the graph so that the degree of all its vertices, modulo those on the boundary are equal to four. We do this in a few steps. First, we note the following simple identity:
 \begin{equation}\label{i1}
\frac{1}{\sqrt{2}}\sum_{S'} e^{i\frac{\pi}{4}SS'}=1.
\end{equation}
 This identity depicted in figure (\ref{UniformUP}), shows that we can add any isolated vertex with an $i\frac{\pi}{4}$ interaction to an existing vertex without any effect except a multiplicative factor of $\sqrt{2}$ which can be taken care of in the final partition function. \\
 To reduce the degree of vertices down to four, we present the following\\

\begin{figure}[t]
\centering
\includegraphics[width=11.5cm,height=2.8cm,angle=0]{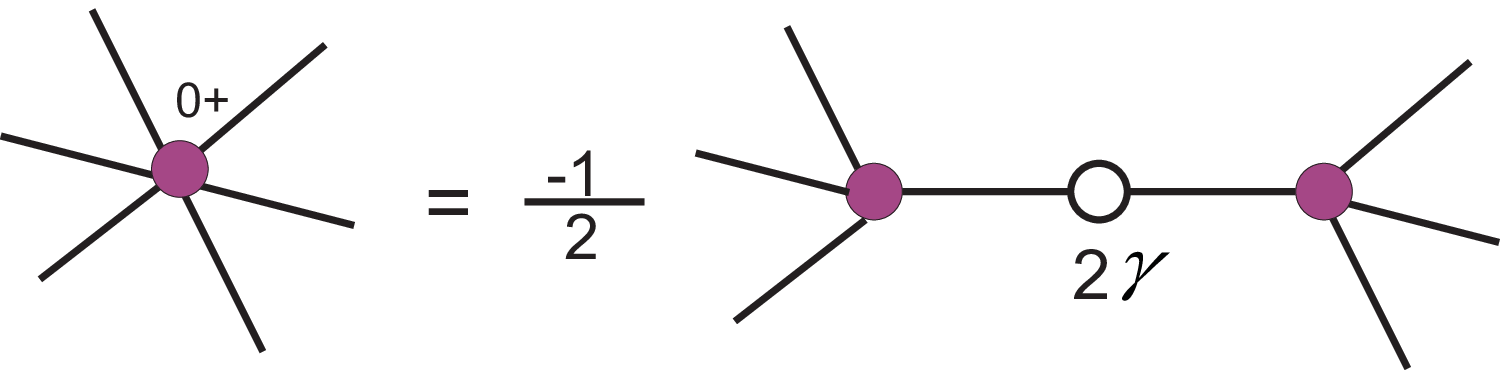}
\caption{(Color Online) Identity (\ref{i2}). The white spin is in magnetic field $i\frac{\pi}{2}$ and is summed over. The result is merging of the two neighboring spins. Reading the diagram from left to right, it shows how a high degree can be split to two vertices of lover degree.
} \label{NewMerging}
\end{figure}

\begin{figure}[t]
\centering
\includegraphics[width=7cm,height=1.3cm,angle=0]{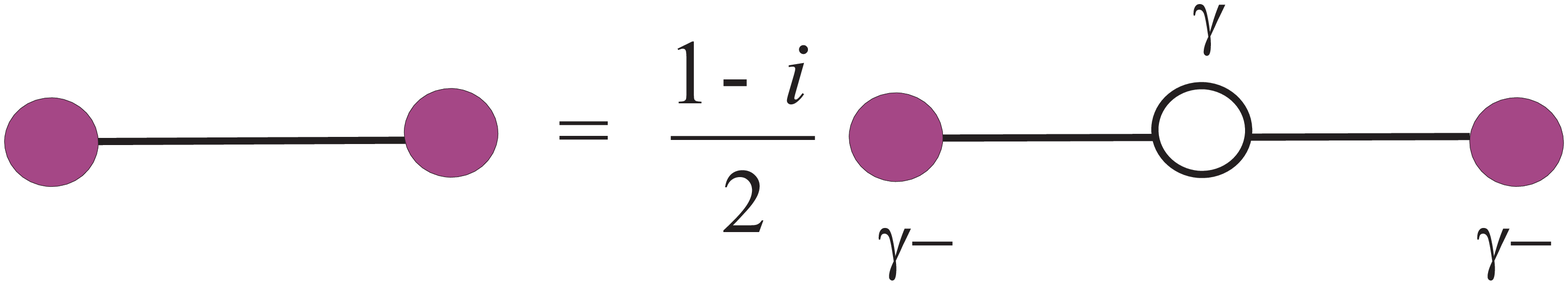}
\caption{(Color Online) Identity (\ref{i3}). Insertion of one vertex on a link: The white spin with a magnetic field $i\frac{\pi}{4}$ can be inserted on a link, reducing the magnetic fields on the endpoints by $i\frac{\pi}{4}$.
} \label{NewLC-1}
\end{figure}

\begin{figure}[t]
\centering
\includegraphics[width=10cm,height=5cm,angle=0]{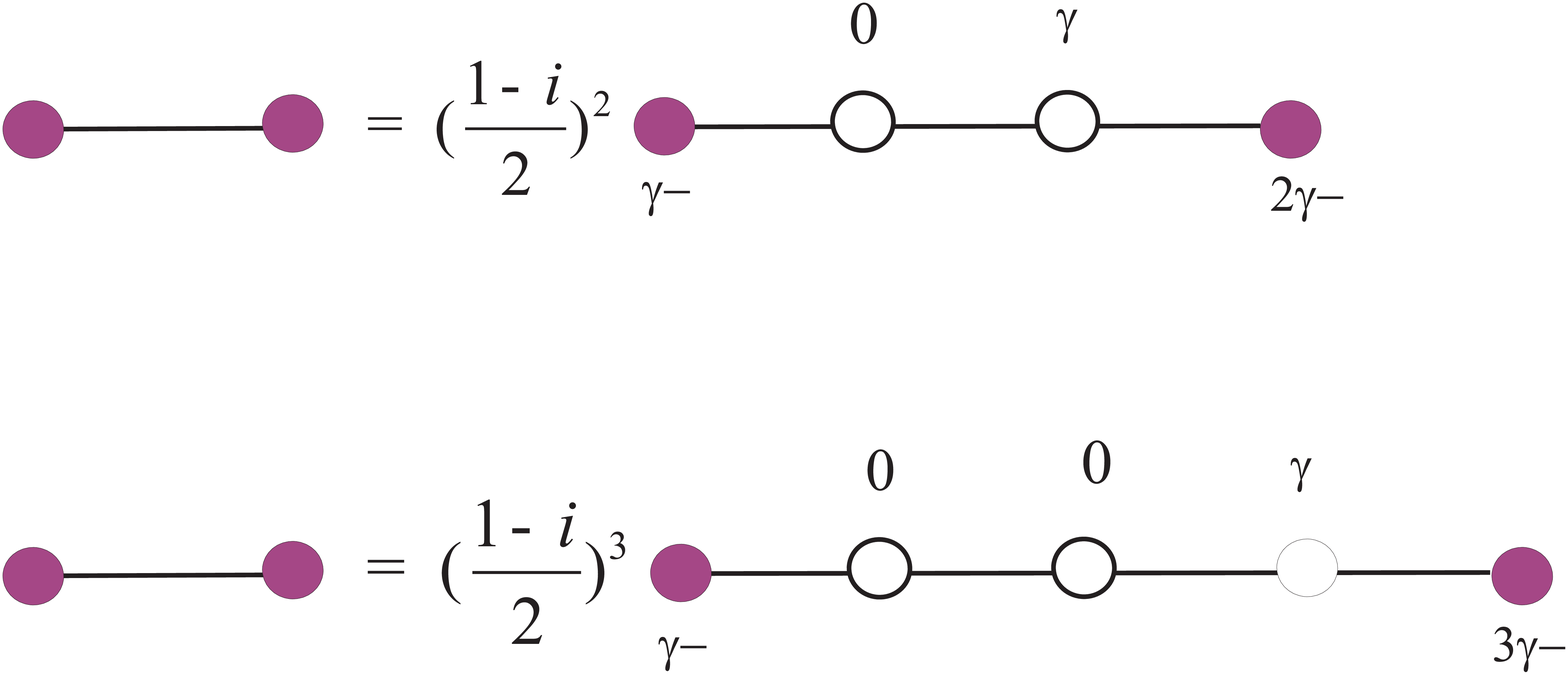}
\caption{(Color Online) Repeating the basic insertion in figure (\ref{NewLC-1}). a) two vertices and (b) three vertices are inserted on the link connecting the black vertices.
} \label{NewLC-1-a}
\end{figure}

 \textbf{Lemma:} For Ising variables (taking values $\pm 1$), the following identity holds:
 \begin{equation}\label{i2}
\delta_{S_1,S_2}= -\frac{1}{2} \sum_{S_0}e^{i\frac{\pi}{2}S_0+i\frac{\pi}{4}S_0(S_1+S_2)}.
\end{equation}
 To verify it, we note that the right hand side is equal to
 \begin{equation}
-\frac{1}{2}\left(ie^{i\frac{\pi}{4}(S_1+S_2)}-ie^{-i\frac{\pi}{4}(S_1+S_2)}\right)=\sin \frac{\pi}{4}(S_1+S_2)=\delta_{S_1, S_2}.
\end{equation}
 This identity depicted graphically in figure (\ref{NewMerging}) allows us to reduce the degree of any vertex down to the value four. Actually as this figure shows, the degree can be reduced to three, however we need degree four for the flattening step which will be explained later.
 Finally we note that we can insert vertices into already existing links and plaquettes. The insertion in a link follows the following identity:
 \begin{equation}\label{i3}
\sum_{S_0}e^{\frac{i\pi}{4}S_0+\frac{i\pi}{4}S_0(S_1+S_2)}=(1+i) e^{\frac{i\pi}{4}(S_1+S_2+S_1S_2)}.
\end{equation}
This identity is diagrammatically shown in figure (\ref{NewLC-1}). It shows that a spin with magnetic field $i\frac{\pi}{4}$ can always be inserted in a link with the effect of reducing the magnetic fields on the end vertices of the link by $i\frac{\pi}{4}$. Repeating this insertion once and twice we arrive at figure (\ref{NewLC-1-a}-a) and (\ref{NewLC-1-a}-b) respectively, with a similar pattern for larger number of insertions.

\begin{figure}[t]
\centering
\includegraphics[width=7cm,height=2.8cm,angle=0]{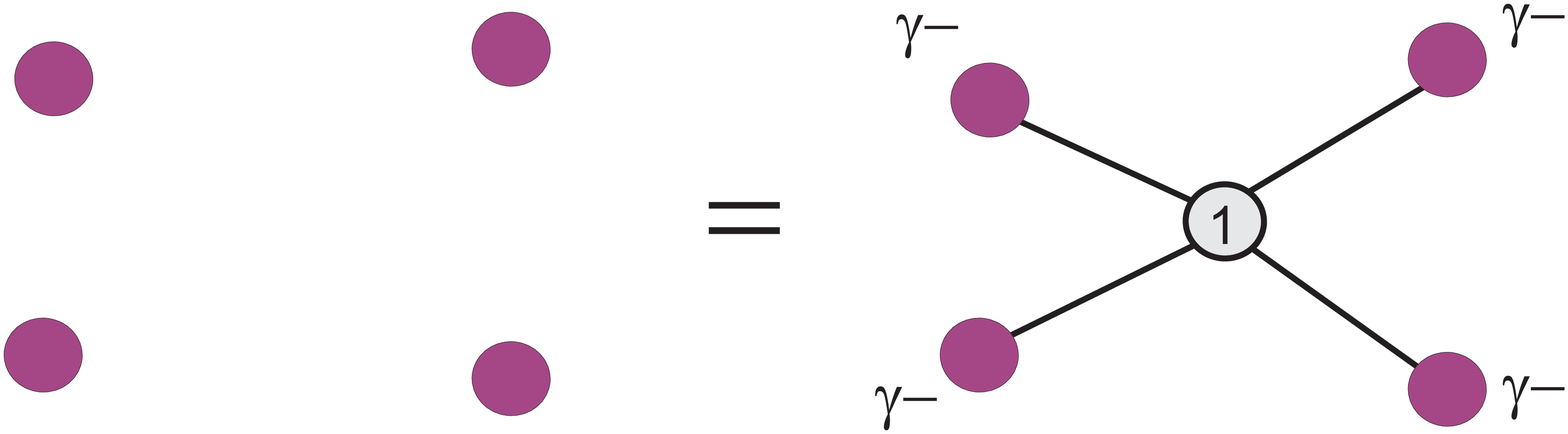}
\caption{(Color Online) Insertion of a spin into a plaquette. The value of the central spin is fixed to $1$ and the magnetic fields of all the spins connected to it are reduced by $i\frac{\pi}{4}$, with no change in the partition function.
} \label{NewLC-1-b}
\end{figure}

 Insertion into a plaquette is achieved simply by fixing the value of that spin to $1$ followed by reducing the magnetic fields on all the spins on the vertices of the plaquette by $i\frac{\pi}{4}$. This is shown in figure (\ref{NewLC-1-b}) and the verification of the identity of the two sides of the figure is trivial.

Therefore with these identities we can uniformize all the degrees of the graph and turn it into a polynomially larger graph with degree four for all vertices.
Some examples for working with these steps are given in the appendix B.

   \subsection{Flattening the Graph}
 The graph obtained by implementing the constraints, may well become a non-planar graph. In this subsection we show that any such graph can be made a planar graph by adding a number of vertices of the same order as the number of vertices in the original graph. Again we present a useful identity in the form of the following \\

 \textbf{Lemma:} Let $\sigma_0$ be a  qubit variable, taking values  $0$ and $1$. Then for any integer $x$, the following identity holds:

 \begin{equation}\label{L3}
\sum_{\sigma_0}e^{i\frac{\pi}{2}(\sigma_0+x)^2}=(1+i).
\end{equation}

 The proof is by simple substitution and by noting that the square of any two consecutive integers are of the form $4k$ and $4k+1$. From this we can then prove the following: \\

 \textbf{Lemma:} Let $\sigma_0$ and $\sigma_i$ be qubit variables.  Then the following identity holds:

 \begin{equation}\label{L4}
\sum_{\sigma_0=0,1}e^{i\frac{\pi}{2}\sigma_0+i\pi\sigma_0(\sum_{j=1}^m\sigma_j)+i\pi\sum_{\la i,j\ra\in K}\sigma_i\sigma_j}=(1+i)e^{-i\frac{\pi}{2}\sum_{j=1}^m\sigma_i+\sum_{\la i,j\ra\in K'}\sigma_i\sigma_j},
\end{equation}
 where $K$ is a subset of the edges which are the neighbors of $\sigma_0$
and $K'$ is the complement of that set.   To prove this we complete the square in (\ref{L4}) and use lemma (\ref{L3}) with $x\equiv \sum_{i=1}^m\sigma_i$. More specifically we write
the left hand side of (\ref{L4}) as
\begin{equation}
{\rm LHS}\ {\rm of} (\ref{L4}) = \sum_{\sigma_0=0,1}e^{i\frac{\pi}{2}(\sigma_0+x)^2} e^{-i\frac{\pi}{2}(\sum_{i=1}^m-i\pi \sum_{i,j\in K'}\sigma_i\sigma_j)}.
\end{equation}
Using lemma (\ref{L3}) and noting that $e^{i\pi Y}=e^{-i\pi Y}$ for any integer, we arrive at (\ref{L4}).  \\

\begin{figure}[t]
\centering
\includegraphics[width=10cm,height=4cm,angle=0]{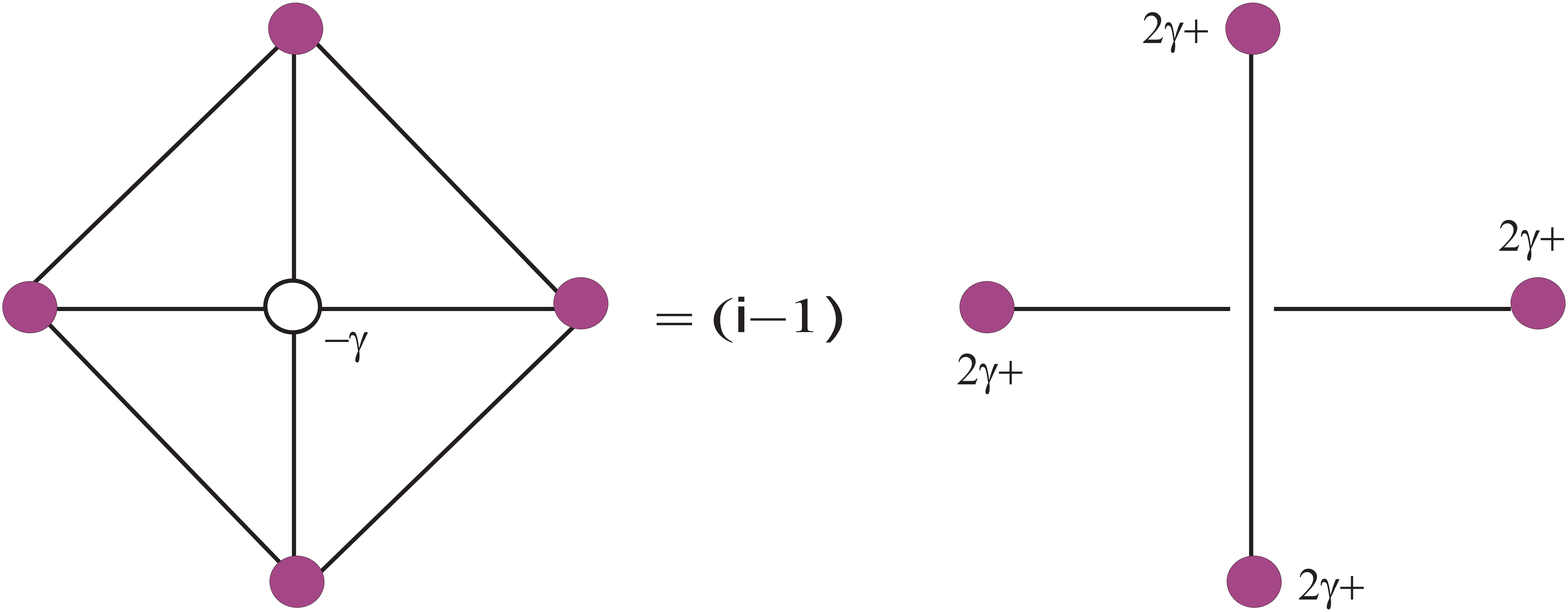}
\caption{(Color Online) Identity (\ref{i4}). The white spin is in magnetic field $-i\frac{\pi}{4}$  and is summed over. The result is complementing the interaction links, as shown, with the addition of magnetic fields $i\frac{\pi}{2}$ to the fields already present in the neighbors. This diagrams allows us to remove all the crossings and flatten the arbitrary graph.
} \label{NewLC-2}
\end{figure}

The lemma (\ref{L4}) which is reminiscent of what is called local complementation rule in the context of measurement-based quantum computation, plays a basic role in flattening every diagram. First note that while the form of the lemma is general in terms of qubit variables $(\sigma=0,1)$, when expressed in terms of Ising variable $(S=1,-1)$, it may not have a simple form, independent of the details of the connections of the graph. However we need this lemma only for a special simple  case. This is the graph of figure (\ref{NewLC-2}). For this simple case the substitution of qubit variables to Ising variables, gives the following simple identity
\begin{equation}\label{i4}
\sum_{S_0}e^{-i\frac{\pi}{4}S_0+i\frac{\pi}{4}S_0(S_1+S_2+S_3+S_4)+i\frac{\pi}{4}(S_1+S_3)(S_2+S_4)}=(i-1)e^{i\frac{\pi}{2}(S_1+S_2+S_3+S_4)+i\frac{\pi}{4}(S_1S_3+S_2S_4)}.
\end{equation}
This identity is shown in graphical form in figure (\ref{NewLC-2}) and it allows us  to remove all the over-crossings and hence flatten all the diagrams.

Therefore by a combination of insertion into the links and faces and removing all the crossings, we can turn the arbitrary graph which is obtained after imposing the constraint into an Ising model on a 2D rectangular lattice. The number of vertices which should be added grows polynomially with the number of spins in the original statistical model, provided that the range of interactions is bounded by some integer independent of the size of the lattice. In this way it is proved that any discrete statistical model is equivalent to a special kind of rectangular 2D Ising model, with the following Hamiltonian:

\begin{equation}\label{i5}
H=\sum_{i}h_i S_i+i\frac{\pi}{4}\sum_{\la i,j\ra}S_iS_j,
\end{equation}
 that is all the neighboring spins interact with each other by the coupling constant $i\frac{\pi}{4}$ and all the original coupling constant and the geometry of the original lattice is reflected into the inhomogeneous magnetic fields on the local spins.

\section{Discussion}
The Ising model in two dimensional rectangular lattices, i.e. 2D Ising model, is a complete model in the sense that any the partition function of any other statistical mechanical model with discrete type variables, is equivalent to a 2D Ising model on a sufficiently large rectangular lattice. This Ising model, whose Hamiltonian is in the form of (\ref{i5}),  is special in the sense that all its coupling constants are equal to the complex value of $i\frac{\pi}{4}$ and all the coupling constants of the original model and all the details of the geometry of the original lattice, and the pattern of interactions  are reflected into the local magnetic fields of the Ising model. This completeness, was first proved \cite{completeness} by using concepts and techniques from quantum information theory and was based on the universality of cluster states. We have now proved this important result,  by using independent and general concepts and methods  which are accessible to a wide audience of researchers across many disciplines.  In fact our method, not only unravels the essence of this completeness from a new perspective, it also provides an algorithmic approach through definite and simple graphical steps for reducing any other model to an Ising model of the above type.
The universal form of the Ising model as shown in Eq. (\ref{i5}), may have far reaching consequences. It shows that all the various  models in different universality classes are different only in the number of sites and the pattern of local magnetic fields on this 2D lattice, and not on the nature of discrete variables, nor on the pattern of interactions. In particular the well-known criteria of dimensionality of the system and symmetry of the order parameter have to be re-interpreted in terms of the complete 2D Ising model.
\section*{Acknowledgements:}

V. K. would like to thank Abdus Salam ICTP for its hospitality, where this paper was finalized.

\section{Appendix A}
In this appendix we provide two simple examples to illustrate the assertion in section (\ref{sec1}) that the discrete variables of any kind of statistical model can always be expressed in terms of Ising variables. The examples are the four-state and three-state Potts models. \\
\begin{figure}[t]
\centering
\includegraphics[width=6.7cm,height=2.7cm,angle=0]{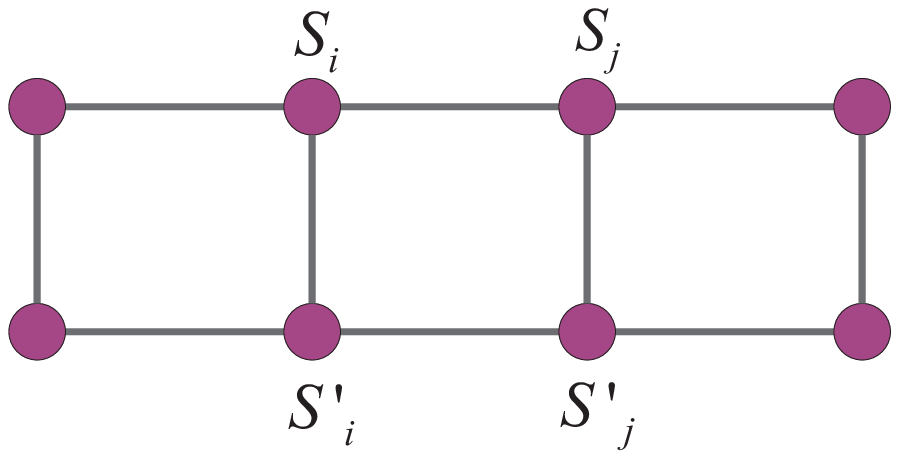}
\caption{(Color Online) Appendix A. The four and three state Potts models can be expressed in terms of Ising variables, see Eqs. (\ref{e1}) and (\ref{e2}).
} \label{B2}
\end{figure}
Consider $1$ dimensional four-state Potts model with the following Hamiltonian:
\begin{equation}
H=-J\sum_{\la i,j \ra} \delta_{q_i,q_j},
\end{equation}
where each $q_i$ takes four different values $\{ 0, 1, 2, 3\}$. We can encode a variable $q_i$ into two Ising variables $(S_i,S'_i)$ and hence turn the one-dimensional chain into a ladder as in figure (\ref{B2}) and use the encoding  $$0\equiv(0,0),\ 1\equiv(0,1),\ 2\equiv(1,0),\ 3\equiv(1,1).$$

A simple calculation then shows that
\begin{equation}\label{e1}
\delta_{q_i,q_j}=\frac{1}{4}(1+S_{i}S_{j}+S'_{i}S'_{j}+S_{i}S_{j}S'_{i}S'_{j}).
\end{equation}

For the three-state Potts model, we use the following encoding  $$0=(0,0)~,~1=(0,1)=(1,0)~,~2=(1,1),$$ where only each Potts value is mapped to the equivalence class of the Ising pair, under symmetry. In this it can be verified that
 \begin{equation}\label{e2}
\delta_{q_{i},q_{j}}=\frac{3}{8}+\frac{1}{8}(-S_{i}S'_{i}-S_jS'_j+S_{i}S_{j}+S'_{i}S'_{j}+S_{i}S'_{j}+S'_{i}S_{j})+\frac{3}{8}S_{i}S'_{i}S_{j}S'_{j}.
\end{equation}
Therefore both models can be expressed in terms of Ising variables.

\section{Appendix B}

In this short appendix we show how the uniformizing rule is to be used via a few simple examples. We present these examples only to show the working of the basic identities. Clearly converting a very complex graph requires many steps to be taken, but the nature of them all is the same as a kind of uniformization and flattening. We consider three simple examples:\\

\subsection*{Example 1:} 
\begin{figure}[t]
\centering
\includegraphics[width=6cm,height=2.5cm,angle=0]{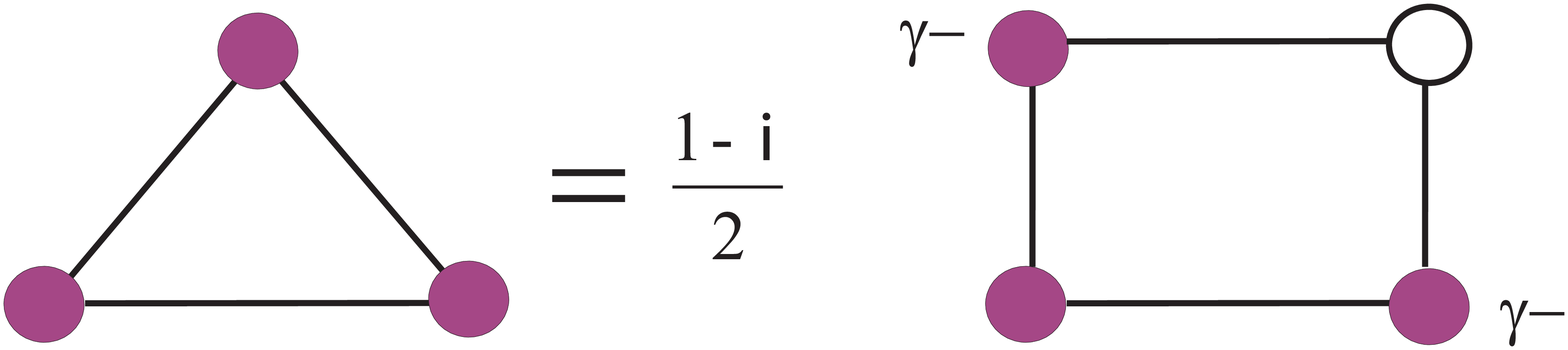}
\caption{(Color Online) Example 1:
Insertion of a vertex on a link, as in (\ref{NewLC-1}), turns a triangle to a rectangle, with appropriate changes in the magnetic fields of the adjacent vertices.
} \label{B}
\end{figure}

Consider a simple triangle as shown in figure (\ref{B}). Insertion into a link, turns this into a rectangular plaquette, the simplest 2D Ising model.

\subsection*{Example 2:}
\begin{figure}[t]
\centering
\includegraphics[width=11cm,height=2.5cm,angle=0]{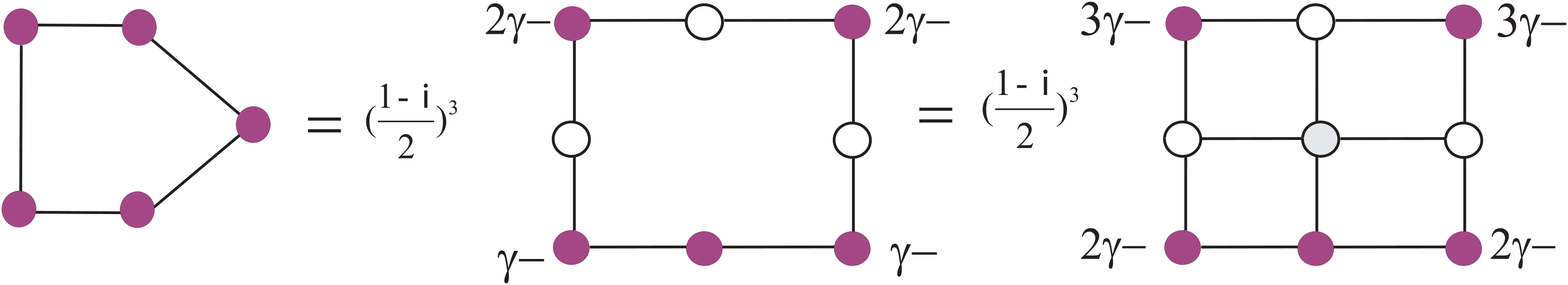}\vspace*{1cm}
\caption{(Color Online) Example 2: Insertion of three vertices on the links of a pentagon, as in (\ref{NewLC-1}), followed by an insertion of a vertex into the plaquette as in (\ref{NewLC-1-b}),  turns a pentagon into a small rectangular lattice.
} \label{C}
\end{figure}

Consider a pentagon as shown in figure (\ref{C}). Three link-insertions followed by a face-insertion  turns this into the rectangular lattice on the right hand side.
\subsection*{Example 3:}
\begin{figure}[t]
\centering
\includegraphics[width=14cm,height=3.5cm,angle=0]{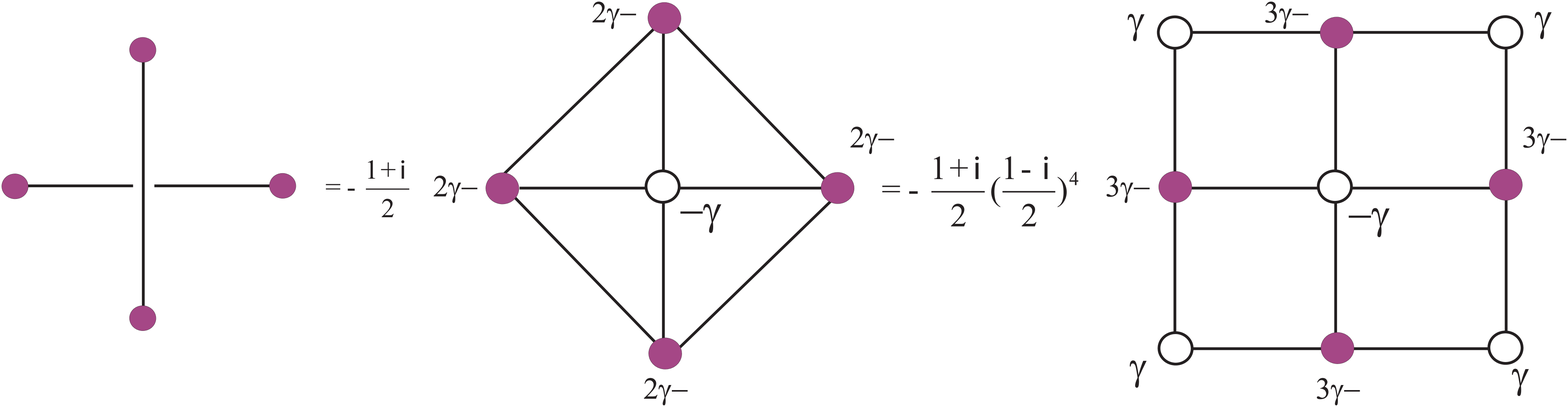}
\caption{(Color Online)  Example 3: By local complementation and insertion of four vertices on the links, we can remove a crossing.
} \label{D}
\end{figure}

Consider a crossing as in figure (\ref{D}). Flattening this according to (\ref{NewLC-2}),  followed by insertions on the four links turns this into a rectangular lattice with four squares.


\begin{thebibliography}{99}
\bibitem{Golden}
Goldenfeld, Lectures On Phase Transitions And The Renormalization Group (1992).
\bibitem{onsager}
L. Onsager, Phys. Rev. 65  117-149 (1944).
\bibitem{kramers}
H. A. Kramers, G. H. Wannier. Phys. Rev. 60 252-262, 263-276 (1941).
\bibitem{kaufman}
B. Kaufman, L. Onsager, Phys. Rev. 76 1232-1243 (1949).
\bibitem{quantum formalism}
M. Van den Nest, W. D\"{u}r and H. J. Briegel, Phys. Rev. Lett. 98,
117207 (2007).
\bibitem{completeness}
Van den Nest, W. D\"{u}r and H. J. Briegel, Phys. Rev. Lett. 100
110501 (2008).

\bibitem{delgado}
De las Cuevas , W. D\"{u}r, H. J. Briegel and Martin-Delgado, Phys. Rev. Lett. 102 230502 (2009).
\bibitem{newj}
G. De las Cuevas, W. D\"{u}r, H. J. Briegel and Martin-Delgado,  New J. Phys. 12 043014 (2010).
\bibitem{u1}
Y. Xu, G. De las Cuevas, W. D\"{u}r, H. J. Briegel, M. A.
Martin-Delgado,  J. Stat. Mech. P02013 (2011).
\bibitem{phi}
V. Karimipour, M. H. Zarei, Phys. Rev. A 85, 032316 (2012)
\bibitem{statistic}
G. De las Cuevas, W. D\"{u}r, M. Van den Nest and H. J. Briegel, J. Stat. Mech. P07001 (2009).
\bibitem{nest}
M. Van den Nest, W. D\"{u}r, R. Raussendorf and H. J. Briegel, Phys. Rev. A80 052334 (2009).
\bibitem{Baxter}
R. J. Baxter, Exactly solved models in statistical mechanics (1982).
\bibitem{nest2}
R. Hubener, M. Van den Nest, W. D\"{u}r and H. J. Briegel, J. Math. Phys. 50 083303 (2009) .
\bibitem{other}
S. Bravyi, R. Raussendorf, Phys. Rev. A 76, 022304 (2007).
\bibitem{vidal}
M. Van den Nest, W. D\"{u}r, G. Vidal, H. J. Briegel, Phys. Rev. A
75, 012337 (2007).
\bibitem{lidar}
D. A. Lidar and O. Biham, Phys. Rev. E 56, 3661 (1997).
\bibitem{lidar1}
F. Verstraete et al., Phys. Rev. Lett. 96, 220601
(2006).
\bibitem{lidar2}
 D. A. Lidar, New J. Phys. 6, 167
(2004).
\bibitem{lidar3}
R. D. Somma, C. D. Batista and
G. Ortiz, Phys. Rev. Lett. 99, 030603 (2007)
\bibitem{lidar4}
 H. Bombin and M.A.
Martin-Delgado, Phys. Rev. A 77, 042322 (2008).
\bibitem{lidar5}
V. Murg, F. Verstraete, I. Cirac, Phys.
Rev. Lett. 95, 057206 (2005).
\bibitem{lidar6}
 F. Verstraete, M.M. Wolf, D. Perez-
Garcia, J. I. Cirac, Phys. Rev. Lett. 96, 220601 (2006).
\bibitem{lidar7}
 J. Geraci, D.A. Lidar, Comm. Math. Phys. 279,
735 (2008).
\bibitem{np}
F. Barahona, J. Phys. A 15, 3241 (1982).
\bibitem{np2}
N. Schuch et al., Phys. Rev. Lett. 98, 140506 (2007).
\bibitem{mqc4}
R. Raussendorf and H. J. Briegel, Phys. Rev. Lett. 86,
5188 (2001).
\bibitem{dmqc1}
D. Gross and J. Eisert, Phys. Rev. Lett. 98, 220503 (2007).
\bibitem{dmqc2}
D. Gross, J. Eisert, N. Schuch, and D. Perez-Garcia, Phys. Rev.
A 76, 052315 (2007).
\bibitem{dmqc3}
D. Gross and J. Eisert, Phys. Rev. A 82, 040303(R) (2010).
\bibitem{dmqc4}
G. K. Brennen and A. Miyake, Phys. Rev. Lett. 101, 010502
(2008).
\bibitem{dmqc5}
X. Chen, B. Zeng, Z. C. Gu, B. Yoshida, and I. L. Chuang, Phys.
Rev. Lett. 102, 220501 (2009).
\bibitem{dmqc6}
C. E. Mora, M. Piani, A.Miyake, M. Van den Nest, W. D\"{u}r, and H. J.
Briegel, Phys. Rev. A 81, 042315 (2010).
\bibitem{dmqc7}
J. M. Cai, W. D\"{u}r, M. Van den Nest, A. Miyake, H. J. Briegel,
Phys. Rev. Lett. 103, 050503 (2009)
\bibitem{dmqc8}
Jianming Cai, Akimasa Miyake, Wolfgang D\"{u}r, Hans J. Briegel,
Phys. Rev. A 82, 052309 (2010)
\bibitem{dmqc9}
Wei-Bo Gao, Xing-Can Yao, Jian-Ming Cai, He Lu, Ping Xu, Tao Yang, Chao-Yang Lu, Yu-Ao Chen, Zeng-Bing Chen, Jian-Wei Pan, Nature Photonics 5, 117 (2011).
\bibitem{mqc}
Nicolas C. Menicucci et al, Phys. Rev. Lett. 97, 110501 (2006).
\bibitem{mqc2}
M. Gu, C. Weedbrook, N. C. Menicucci, T. C. Ralph, and P.
vanLoock, Phys. Rev. A. 79 , 062318 (2009).
\bibitem{mqc3}
Dan Browne and Hans Briegel, arXiv:0603226v2 (2006).
\bibitem{mqc5}
S. D. Bartlett, B. C. Sanders, S. L. Braunstein, and K. Nemoto,
Phys. Rev. lett. 88 , 097904 (2002).
\bibitem{mqc6}
N. C. Menicucci, S. T. Flammia, and O. Pfister, Phys. Rev. Lett.
101, 130501 (2008).
\bibitem{mqc7}
S. T. Flammia, N. C. Menicucci, and O. Pfister, J. Phys. B 42,
114009 (2009).
\bibitem{color}
M. Delgado, H. Bombin, Phys. Rev. Lett. 97, 180501
(2006)
\bibitem{three}
D. W. Wood and H. P. Griffiths, J. Phys. C 5, L253 (1972).
\bibitem{three2}
R. J. Baxter and F. Y. Wu, Phys. Rev. Lett. 31, 1294 (1973).
\end{thebibliography}
\end{document}